\documentclass[twocolumn,aps,prb]{revtex4-2}

\usepackage{amsfonts}
\usepackage{amsmath}
\usepackage{amssymb}
\usepackage[pdftex]{graphicx}
\usepackage{dcolumn}
\usepackage{bm}
\usepackage{braket}
\usepackage{color}
\usepackage{here}
\usepackage{mathrsfs}
\usepackage{mathtools}
\usepackage[normalem]{ulem}

\newcommand{\bkakko}[1]{\left[#1\right]}

\newcommand{\us}{\uparrow}
\newcommand{\ds}{\downarrow}

\begin{document}

\title{Thermal conductance at the superradiant phase transition in the quantum Rabi model}

\author{Tsuyoshi Yamamoto}
\email{Contact author: tsuyoshi.yamamoto@cea.fr}
\thanks{\protect\\Present address: Univ. Grenoble Alpes, CEA, Grenoble INP, IRIG, PHELIQS, 38000 Grenoble, France}
\author{Yasuhiro Tokura}
\affiliation{Institute of Pure and Applied Sciences, University of Tsukuba, Tsukuba, Ibaraki 305-8577, Japan}
\date{\today}

\begin{abstract}
The quantum Rabi model exhibits a superradiant phase transition when the coupling becomes strong, even though it involves only two components: a two-level atom and a single bosonic mode.
This phase transition is referred to as a finite-component quantum phase transition, in contrast to conventional phase transitions in many-body systems.
In this Letter, we investigate heat transport across an atom embedded in bosonic modes, modeled by the quantum Rabi model, between two thermal baths.
We found a manifestation of the superradiant phase transition in the thermal conductance, which represents the linear response to a temperature bias.
Our Letter can be helpful for the development of quantum heat devices utilizing controllable finite-component quantum phase transitions.
\end{abstract}

\pacs{Valid PACS appear here}

\maketitle

{\it Introduction.}
The interactions between light and matter are ubiquitous phenomena in nature and have been extensively studied in a wide variety of physical systems from those of quantum optics~\cite{Scully_text, Vedral_text} to those of molecular physics~\cite{Schwartz2011, George2016, Benz2016}.
The quantum Rabi model~\cite{Braak2010}, consisting of a single bosonic mode with frequency $\Omega$ and a two-level atom with frequency $\Delta$, is the simplest model that captures the essence of light-matter interactions.
It serves as not only a fundamental theoretical model but also a platform to experimentally explore the extent to which light and matter can interact strongly~\cite{Kockum2019, Forn-Diaz2019, Qin2024}.
Recent advancements in artificial systems, such as superconducting circuits and ion trap systems, have made it possible to conduct experiments in the ultrastrong~\cite{Anappara2009, Forn-Diaz2010, Forn-Diaz2017} ($\lambda/\Omega\gtrsim0.1$) and deep-strong coupling regimes~\cite{Bayer2017, Yoshihara2017, Yoshihara2018} ($\lambda/\Omega\gtrsim1$), where $\lambda$ denotes the coupling strength.
In these regimes, strong light-matter interactions manifest novel physical phenomena as well as have quantum-information applications~\cite{Nielsen_text, Nataf2011, Romero2012, Ofek2016}.

Quantum phase transitions are hallmarks of phenomena induced by strong interactions.
Conventional quantum phase transitions are associated with many-body systems in the thermodynamic limit~\cite{Sachdev_text}.
However, even in the quantum Rabi model, which involves only two constituents, a quantum phase transition can occur in the classical oscillator limit, called a finite-component quantum phase transition~\cite{Hwang2010, Ashhab2010, Ashhab2013, Hwang2015, Shen2021, Grimaudo2024, Zheng2024, Puebla2020, Hwang2016, Garbe2020, Zhu2020}.
When the coupling strength exceeds a critical value $\lambda_c=\sqrt{\Omega\Delta}/2$, $\mathbb{Z}_2$ parity symmetry is spontaneously broken, resulting in a superradiant phase transition from the zero coherence of the bosonic mode (normal phase) to a finite coherence (superradiant phase).
Recently, the superradiant phase transition in the quantum Rabi model was experimentally observed in a trapped ion~\cite{Cai2021} and in nuclear magnetic resonance~\cite{Chen2021}, where the average bosonic occupation number, a static order parameter, was observed to change abruptly across the critical point.
These experiments have focused attention on quantum phase transitions in controllable finite-component systems, not only from a fundamental statistical mechanics perspective but also for their potential applications to quantum devices.

The influence of quantum phase transitions extends beyond equilibrium properties; they play a crucial role in non-equilibrium transport phenomena such as quantum heat transport~\cite{Giazotto2006, Pekola2021}.
For instance, in quantum heat transport through a two-level system, the linear thermal conductance exhibits a nontrivial power-law temperature dependence in the quantum critical regime~\cite{Yamamoto2018PRB}.
While quantum heat transport in the quantum Rabi model (or the Jaynes-Cummings model~\cite{Jaynes1963}) has been investigated~\cite{Yamamoto2021, Xu2021, Wang2021, Chen2022, Magazzu2024}, the impact of the superradiant phase transition on heat transport remains largely unexplored.

Moreover, a deeper understanding of quantum heat transport would have significant implications for the development of quantum heat devices for managing heat in nanoscale systems.
Recently, quantum heat valves and thermal rectifiers have been experimentally demonstrated in the weak coupling regime~\cite{Ronzani2018, Senior2020}, and quantum heat transport has also begun to be observed in the ultra-strong coupling regime~\cite{Upadhyay2024}.
Theoretical studies have proposed leveraging collective phenomena involving $N$ atoms to enhance the performance of quantum heat devices~\cite{Vogl2011,  Kamimura2022, Kolisnyk2023, Andolina2024}.
In this Letter, we investigate the enhancement of heat current involving a single atom.
The abrupt changes in the ground state induced by a finite-component quantum phase transition under strong coupling are expected to pave the way for novel quantum heat devices.

\begin{figure}[tb]
    \centering
    \includegraphics[width=1\linewidth]{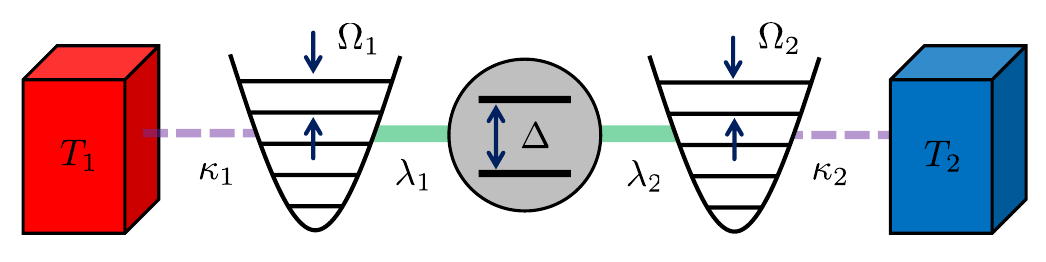}
    \caption{Schematic diagram of the two-mode quantum Rabi model with dissipation through two thermal baths. The atom is indirectly coupled to the thermal baths via the bosonic mode.}
    \label{fig:setup}
\end{figure}

{\it Hamiltonian.}
Here, we consider a two-level atom in contact with two thermal baths via single bosonic modes (see Fig.~\ref{fig:setup}), the Hamiltonian of which is given by $H=H_{\rm S}+H_{\rm B}+V$.
The composite system, the atom embedded between the two bosonic modes, is described by the two-mode quantum Rabi model,
\begin{align}
H_{\rm S}=\sum_{r=1,2}\hbar\Omega_ra_r^\dagger a_r+\frac{\hbar\Delta}{2}\sigma_z-\sum_{r=1,2}\hbar\lambda_r(a_r+a^\dagger_r)\sigma_x,
\end{align}
where $a_r$ ($a_r^\dagger$) is a bosonic annihilation (creation) operator of the bosonic mode $r$ with the resonant frequency $\Omega_r$, $\sigma_{i}~(i=x,y,z)$ is the Pauli operator acting on the atom with energy splitting $\hbar\Delta$ so that $\sigma_z=\ket{\us}\bra{\us}-\ket{\ds}\bra{\ds}$, and $\lambda_r$ represents the coupling strength between the bosonic mode and the atom.
The thermal bath is modeled as a collection of harmonic oscillators, $H_{\rm B}=\sum_{r}H_{{\rm B},r}=\sum_{r,k}\hbar\omega_{rk}b_{rk}^\dagger b_{rk}$, where $b_{rk}$ is a bosonic operator of the mode $k$ in the thermal bath $r$ with natural frequency $\omega_{rk}$.
The interaction between the bosonic modes and the thermal baths is represented by $V=\sum_r V_r=\sum_{r,k}\hbar\eta_{rk}(a_r+a_r^\dagger)(b_{rk}+b_{rk}^\dagger)$.
Here, we assume Ohmic dissipation, which is characterized by a spectral density, $I_r(\omega)\equiv\sum_k \eta_{rk}^2\delta(\omega-\omega_{rk})=2\kappa_r\omega$ with a dimensionless coupling strength $\kappa_r$~\cite{Leggett1987, Weiss_text}.
For a weak dissipation ($\kappa_r\ll1$), the rotating-wave approximation is applicable, and thus, the counter-rotating terms are negligible, i.e., $V_r\approx\sum_k\hbar\eta_{rk}(a_rb_{rk}^\dagger+a_r^\dagger b_{rk})$.
In this Letter, we consider the symmetric case, $\Omega=\Omega_1=\Omega_2$, $\lambda=\lambda_1=\lambda_2$, $\eta_{k}=\eta_{1k}=\eta_{2k}$, and $\omega_k=\omega_{1k}=\omega_{2k}$, resulting in $\kappa=\kappa_1=\kappa_2$.

{\it Thermal conductance.}
The heat current flowing out of thermal bath 1 is defined as $J_1(t)\equiv-\braket{\dot{H}_{{\rm B},1}(t)}$, where $\braket{\cdot}$ denotes the quantum mechanical average with the total Hamiltonian and $H_{{\rm B},1}(t)=e^{iHt/\hbar}H_{{\rm B},1}e^{-iHt/\hbar}$.
Using the Keldysh formalism, we obtain the steady-state heat current $J=J_1=-J_2$ as (see Supplemental Material~\footnote{See Supplemental Material for a detailed derivation of the steady-state heat current.})
\begin{align}
\label{eq:J}
J&=-\hbar^2{\rm Im}\int_0^\infty d\omega~\omega I(\omega)\Big[G^{\rm R}_1(\omega)n_1(\omega)-G^{\rm R}_2(\omega)n_2(\omega) \nonumber\\
&\qquad\qquad\qquad\qquad\qquad\qquad
-\frac{G^<_1(\omega)-G^<_2(\omega)}{2}\Big],
\end{align}
where $G^{\rm R}_r(t,t')=(i\hbar)^{-1}\theta(t-t')\braket{[a_r(t),a_r^\dagger(t')]}$ and $G^<_r(t,t')=(i\hbar)^{-1}\braket{a^\dagger_r(t')a_r(t)}$ are the retarded and lesser Green functions of the bosonic modes, respectively, and $n_r(\omega)=1/(e^{\beta_r\hbar\omega}-1)$ is the Bose-Einstein distribution function of the thermal bath $r$ with temperature $T_r=1/(k_{\rm B}\beta_r)$.

When the temperature bias is small $\Delta T\ll T$ where $T_{1/2}=T\pm \Delta T/2$, its linear response is the thermal conductance, defined as $G(T)\equiv\lim_{\Delta T\to 0}J/\Delta T$.
Using Eq.~\eqref{eq:J}, we can write the thermal conductance as
\begin{align}
G(T)=-k_{\rm B}\hbar\kappa\int d\omega~\omega\sum_{r}{\rm Im}[G^{\rm R}_r(\omega)]\bkakko{\frac{\beta\hbar\omega/2}{\sinh(\beta\hbar\omega/2)}}^2.
\end{align}

Introducing the symmetric and antisymmetric modes of the bosonic mode and the thermal bath, $a_\pm=(a_1\pm a_2)/\sqrt{2}$ and $b_{\pm,k}=(b_{1k}\pm b_{2k})/\sqrt{2}$, respectively, the total Hamiltonian is separated into two modes as $H=H_++H_-$, where $H_\pm=H_{{\rm S},\pm}+H_{{\rm B},\pm}+V_\pm$.
The Hamiltonians of the thermal bath and dissipation retain the form they had in the original frame, i.e., $H_{{\rm B},\pm}=\sum_k\hbar\omega_kb_{\pm,k}^\dagger b_{\pm,k}$ and $V_\pm=\sum_k\hbar\eta_k(a_\pm b_{\pm,k}^\dagger+a_\pm^\dagger b_{\pm,k})$.
For the symmetric mode, the Hamiltonian of the composite system reduces to the conventional quantum Rabi model,
\begin{align}
\label{eq:HR}
H_{{\rm S},+}=\hbar\Omega a_+^\dagger a_++\frac{\hbar\Delta}{2}\sigma_z-\sqrt{2}\hbar\lambda(a_++a_+^\dagger)\sigma_x\equiv H_{\rm R},
\end{align}
while for the antisymmetric mode, the atom is decoupled from the bosonic mode, as $H_{{\rm S},-}=\hbar\Omega a_-^\dagger a_-$.
The thermal conductance is the sum of those for each mode, $G=G_++G_-$, where
\begin{align}
G_\pm(T)=-k_{\rm B}\hbar\kappa\int d\omega~\omega{\rm Im}[G_\pm^{\rm R}(\omega)]\bkakko{\frac{\beta\hbar\omega/2}{\sinh(\beta\hbar\omega/2)}}^2,
\end{align}
where $G_\pm^{\rm R}(t,t')=(i\hbar)^{-1}\theta(t-t')\braket{[a_\pm(t),a_\pm^\dagger(t')]}$.
For a weak dissipation ($\kappa\ll1$), using perturbation theory on the dissipation $V_\pm$, we obtain the full retarded Green function $[G_\pm^{\rm R}(\omega)]^{-1}=[g_\pm^{\rm R}(\omega)]^{-1}-\Sigma(\omega)$, where $g_\pm^{\rm R}(\omega)$ is the retarded Green function of the isolated composite system $H_{{\rm S},\pm}$ and $\Sigma(\omega)=\hbar\sum_{k}\eta_k^2\mathcal{P}\frac{1}{\omega-\omega_k}-i\pi\hbar I(\omega)$ is the self-energy (see Supplemental Material~\footnote{See Supplemental Material explaining perturbation theory on the dissipation.}).
Note that, to evaluate the thermal conductance, it is sufficient to calculate only the retarded Green function of the isolated composite system $g_\pm^{\rm R}(\omega)$.

{\it Antisymmetric mode.}
For the antisymmetric mode, the isolated composite Hamiltonian is a single bosonic mode.
We can calculate the isolated retarded Green function as $g_-^{\rm R}(\omega)=1/(\hbar\omega-\hbar\Omega+i0_+)$; accordingly, the full retarded Green function reads as $G_-^{\rm R}(\omega)=1/[\hbar\omega-\hbar\Omega_\star+i\hbar\pi I(\omega)]$, where $\Omega_\star=\Omega-\sum_k\eta_k^2\mathcal{P}\frac{1}{\omega-\omega_k}$ is the renormalized resonant frequency.
For a weak dissipation, the frequency shift is negligible, $\Omega_\star\approx\Omega$.
Therefore, we obtain the antisymmetric thermal conductance,
\begin{align}
\label{eq:G_asym}
G_-(T)=16k_{\rm B}\Omega\pi\kappa^2\frac{k_{\rm B}T}{\hbar\Omega}\int dx~K_-(x)\frac{x^4}{\sinh^2 x}.
\end{align}
where $1/K_-(x)=(2x-\beta\hbar\Omega)^2+(4\pi\kappa x)^2$.
At low temperatures ($k_{\rm B}T\ll\hbar\Omega$), we can approximate $K_{-}(x)$ as a constant, $K_{-}(x)\approx1/(\beta\hbar\Omega)^{2}$; then, the thermal conductance shows asymptotes $G_-(T)\sim\frac{8}{15}k_{\rm B}\Omega\pi^5\kappa^2(k_{\rm B}T/\hbar\Omega)^3$.
The cubic temperature dependence of the thermal conductance has also been observed in quantum heat transport through a two-level system at low temperatures, where heat is transferred via virtual excitations~\cite{Ruokola2011, Saito2013}.
For the intermediate temperature range ($k_{\rm B}T\sim\hbar\Omega$), $K_{-}(x)$ can be approximated as a Lorentzian with mean $x_0=\beta\hbar\Omega/2$ and width $\gamma=\pi\kappa\beta\hbar\Omega$, i.e., $K_{-}(x)\approx(4\kappa\beta\hbar\Omega)^{-1}\times\frac{1}{\pi}\frac{\gamma}{(x-x_0)^2+\gamma^2}$.
For a weak dissipation, the Lorentzian can be approximated as a delta function.
Thus, the thermal conductance takes $G_-(T)\sim k_{\rm B}\Omega\pi\kappa(\beta\hbar\Omega/2)^2/\sinh^2(\beta\hbar\Omega/2)$, which corresponds to heat transport by a sequential tunneling process~\cite{Segal2005, Yamamoto2018}.
For $k_{\rm B}T\gtrsim\hbar\Omega$, the thermal conductance becomes temperature-independent, $G_-(T)\approx k_{\rm B}\Omega\pi\kappa$.
As the temperature decreases to $k_{\rm B}T\lesssim\hbar\Omega$, the thermal conductance exhibits the Schottky-type temperature dependence, $G_-(T)\approx k_{\rm B}\Omega\pi\kappa(\hbar\Omega/k_{\rm B}T)^2e^{-\hbar\Omega/k_{\rm B}T}$, and heat transport is exponentially suppressed.
At high temperatures ($k_{\rm B}T\gg\hbar\Omega$), since only the high-frequency tail of $K_{-}(x)\sim 1/(2x)^2$ is predominant, the asymptote of the thermal conductance is $G_-(T)\sim\frac{2}{3}k_{\rm B}\Omega\pi^3\kappa^2(k_{\rm B}T/\hbar\Omega)$.
This linear temperature dependence arises from the frequency-independent transmission through the atom in contact with the thermal baths.

{\it Classical~oscillator~limit.}
The quantum Rabi model~\eqref{eq:HR} exhibits a superradiant phase transition between the normal phase ($g=2\sqrt{2}\lambda/\sqrt{\Delta\Omega}>1$) and superradiant phase ($g<1$) in the classical oscillator limit ($\Omega/\Delta\to0$)~\cite{Hwang2015}.
For the time being, we will omit the subscript ``+'' indicating the symmetric mode.
For $\Omega\ll\Delta$, we can obtain the diagonalized form of the quantum Rabi model as
\begin{align}
\label{eq:Hn}
H_n=S^\dagger\bra{\uparrow}U^\dagger H_{\rm R}U\ket{\uparrow}S
\approx\epsilon_n(g)a^\dagger a+\epsilon^0_n(g),
\end{align}
where $U=e^{i\sqrt{2}(\lambda/\Omega)(a+a^\dagger)\sigma_y}$ and $S=e^{-r(g)(a^{\dagger2}-a^2)/2}$ are the Schrieffer-Wolf operator and the squeezing operator with $r(g)=-(1/4)\ln(1-g^2)$, respectively.
The excitation energy is $\epsilon_n(g)=\hbar\Omega\sqrt{1-g^2}$ and the ground-state energy is $\epsilon_n^0(g)=(\epsilon_n(g)-\hbar\Omega-\hbar\Delta)/2$~\cite{Hwang2015}.
The expression for the excitation energy indicates that the above procedure fails for $g>1$. 
For $g>1$, after first applying the displacement operator $D_\pm=e^{\pm\alpha(a^\dagger-a)}$ with $\alpha=\sqrt{\Delta(g^2-g^{-2})/(4\Omega)}$, the quantum Rabi model can be written as the diagonalized form in a similar way to the case of $g<1$ using $\tilde{\lambda}=\sqrt{\Omega\Delta}/(2g)$, $\tilde{\Delta}=g^2\Delta$, and $\tilde{g}=g^{-2}$ instead of $\lambda$, $\Delta$, and $g$,
\begin{align}
\label{eq:Hs}
H_{s,\pm}
=\tilde{S}^\dagger \bra{\tilde{\ds}_\pm}\tilde{U}_\pm^\dagger D_\pm^\dagger H_{\rm R} D_\pm \tilde{U}_\pm\ket{\tilde{\ds}_\pm}\tilde{S}
\approx\epsilon_s(g)a^\dagger a+\epsilon_s^0(g),
\end{align}
where $\ket{\tilde{\ds}_\pm}=(\pm\sqrt{1-g^{-2}}\ket{\us}+\sqrt{1+g^{-2}}\ket{\ds})/\sqrt{2}$.
The excitation energy is $\epsilon_s(g)=\hbar\Omega\sqrt{1-g^{-4}}$ and the ground-state energy is $\epsilon_s^0(g)=(\epsilon_s(g)-\hbar\Omega-\hbar\Delta(g^2+g^{-2})/2)/2$.
The low-energy effective Hamiltonian is independent of the sign of the displacement.
This indicates that it is two-fold degenerate for $g>1$.
Note that the diagonalized Hamiltonian for each phase reproduces the results from Ref.~\cite{Hwang2015}.

These low-energy effective Hamiltonians are exact in the classical oscillator limit.
At the boundary, $g=1$, the gap closes, which indicates a quantum phase transition.
They allow us to calculate the coherence of the bosonic mode: $\braket{a}_{\rm R}=0$ (normal phase) for $g<1$ and $\braket{a}_{\rm R}=\pm\alpha$ (superradiant phase) for $g>1$, where $\braket{\cdot}_{\rm R}$ denotes the quantum mechanical average with the quantum Rabi model.
The finite coherence is characteristic of the superradiant phase transition~\cite{Hepp1973, Hwang2015}.

{\it Normal~phase.}
The low-energy effective Hamiltonian~\eqref{eq:Hn} is quadratic.
Thus, the retarded Green function of the quantum Rabi model can be calculated as
\begin{align}
g_+^{\rm R}(\omega)=\frac{\cosh^2[r(g)]}{\hbar\omega-\epsilon_n(g)+i0_+}-\frac{\sinh^2[r(g)]}{\hbar\omega+\epsilon_n(g)+i0_+},
\end{align}
where we have used the transformed annihilation operator of the bosonic mode, $S^\dagger \bra{\ds}U^\dagger a U\ket{\ds}S=\cosh[r(g)]a^\dagger+\sinh[r(g)]a$.
The thermal conductance is obtained as
\begin{align}
\label{eq:G_np}
G_+^n(T,g)=16k_{\rm B}\Omega\pi\kappa^2\frac{k_{\rm B}T}{\hbar\Omega}\int dx~K_+^n(x)\frac{x^4}{\sinh^2x},
\end{align}
where $1/K_{+}^{n}(x)=[((2x)^2-(\beta\hbar\Omega)^2(1-g^2))/(2x+\beta\hbar\Omega(1-g^2/2))]^2+(4\pi\kappa x)^2$.
Note that the thermal conductance is independent of the atomic energy $\hbar\Delta$ because the atomic state is projected to the $\ket{\ds}$ subspace in the low-energy effective Hamiltonian.
In the decoupling limit ($g\to0$), one can confirm that the thermal conductance of the symmetric mode is equal to the thermal conductance of the antisymmetric mode, i.e., $G_{+}^n(T,g\to0)=G_-(T)$.

At low temperature ($k_{\rm B}T\ll \epsilon_n(g)$), the approximation $K_{+}^n(x)\approx(2\beta\hbar\Omega)^{-2}[1+(\hbar\Omega/\epsilon_n(g))^2]^2$ leads to an asymptotic form of the thermal conductance: $G_+^n(T,g)\sim\frac{2}{15}k_{\rm B}\Omega \pi^5\kappa^2[1+(\hbar\Omega/\epsilon_n(g))^2]^2(k_{\rm B}T/\hbar\Omega)^3$.
This asymptote exhibits significant enhancement near $g=1$ as $G_+^n(g\to1_-)\propto(1-g^2)^{-2}$; this is a manifestation of the superradiant phase transition, accompanied by the closing of the excitation gap.
As $g$ approaches the critical point, the occupation number of the bosonic mode associated with the atom increases, along with its fluctuations, resulting in enhanced heat transport mediated by the bosonic mode.
For the intermediate temperature range ($k_{\rm B}T\sim\epsilon_n(g)$), the function $K_+^n(x)$ can be approximated as a Lorentzian with mean $\beta\epsilon_n(g)/2$ and width $\pi\kappa\beta(\epsilon_n(g)+\hbar\Omega)^2/4\hbar\Omega$.
For a weak dissipation ($\kappa\ll1$), the thermal conductance is $G_+^n(T,g)\sim\frac{1}{4} k_{\rm B}\Omega\pi\kappa(\epsilon_n(g)+\hbar\Omega)^2/(\hbar\Omega)(\beta\epsilon_n(g)/2)^2/\sinh^2(\beta\epsilon_n(g)/2)$.
In a similar way to the antisymmetric mode, the thermal conductance becomes temperature independent for $k_{\rm B}T\gtrsim\epsilon_n(g)$ and shows the Schottky-type temperature dependence for $k_{\rm B}T\lesssim\epsilon_n(g)$ although in the symmetric mode, it depends on $g$.
At high temperatures ($k_{\rm B}T\gg\epsilon_n(g)$), only the tail of $K_+^n(x)\approx1/(2x)^2$ contributes to the thermal conductance, and the high-temperature asymptote of the thermal conductance is the same as in the antisymmetric mode, $G_+^n(T,g)\sim \frac{2}{3}k_{\rm B}\Omega\pi^3\kappa^2(k_{\rm B}T/\hbar\Omega)$.
This is because the small energy structure, $\sim\hbar\Omega$, is smeared out by thermal fluctuations.

\begin{figure}[tb]
    \centering
    \includegraphics[width=1\linewidth]{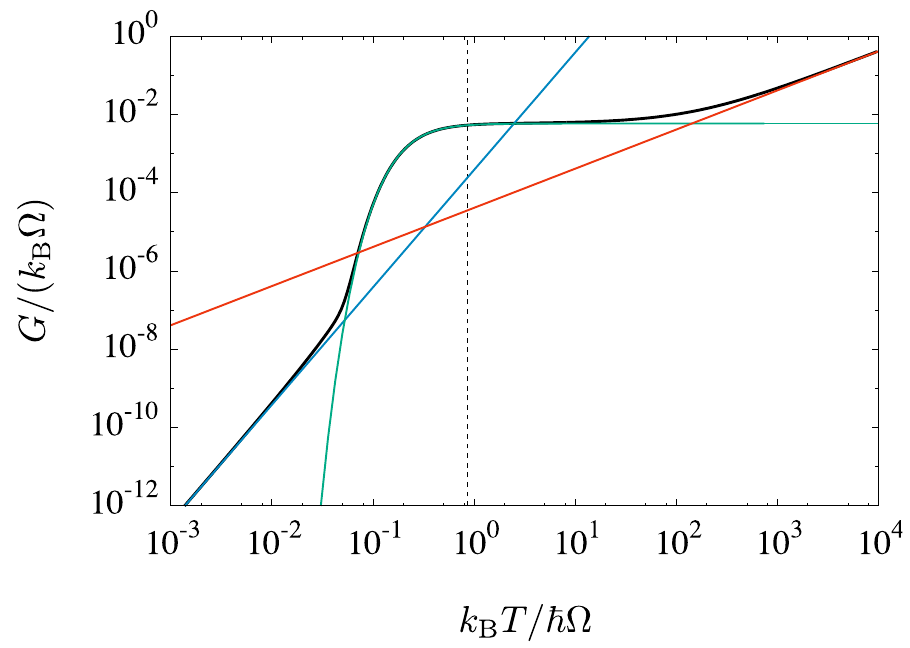}
    \caption{Temperature dependence of thermal conductance for $g=0.5$ and $\kappa=0.001$, calculated using Eqs.~\eqref{eq:G_asym} and \eqref{eq:G_np} The colored (blue, green, and red) lines represent the asymptotes for low, intermediate, and high temperatures, respectively. The dashed line corresponds to the excitation energy $\epsilon_n(g)$.}
    \label{fig:GT}
\end{figure}

Figure~\ref{fig:GT} shows the full temperature dependence of the thermal conductance, $G(T,g)=G_-(T)+G_+^n(T,g)$, for $g=0.5$ and $\kappa=0.001$ with the asymptotes for each temperature range.
Since $\epsilon_n(g=0.5)/\hbar\Omega\approx0.866$, the low-temperature asymptote, which is proportional to $T^3$, is in good agreement with the thermal conductance for $k_{\rm B}T/\epsilon_n(g)\lesssim0.05$, and the high-temperature asymptote, linear to $T$, matches it for $k_{\rm B}T/\epsilon_n(g)\gtrsim500$.
Between these two asymptotic regions, $k_{\rm B}T\sim\epsilon_n(g)$, the thermal conductance is well described by the intermediate-temperature asymptote; it remains temperature independent for $k_{\rm B}T\gtrsim\epsilon_n(g)$ and decays exponentially as the temperature decreases.

{\it Superradiant~phase.}
Since the low-energy effective Hamiltonian~\eqref{eq:Hs} is also a quadratic similarly to the normal phase, we can utilize the results for the thermal conductance in the normal phase.
However, the transformed annihilation operator of the bosonic mode is $\tilde{S}^\dagger \bra{\tilde{\ds}_\pm}\tilde{U}_\pm^\dagger D_\pm^\dagger a D_\pm\tilde{U}_\pm \ket{\tilde{\ds}_\pm}\tilde{S}
=\cosh[r(\tilde{g})]a+\sinh[r(\tilde{g})]a^\dagger\pm\alpha$, different from the normal phase.
Here, the characteristic of the superradiant phase is the additional constant term $\pm\alpha$ resulting in the finite coherence of the bosonic mode, $\braket{a}_{\rm R}=\pm\alpha$.
However, this constant term does not appear in the dynamical response function describing fluctuations.
Therefore, the thermal conductance has the same expression as in the normal phase with $\tilde{g}=g^{-2}$,
\begin{align}
\label{eq:Gsp}
G_+^s(T,g)=2G_+^n(T,g^{-2}),
\end{align}
where the coefficient $2$ arises from the two-fold degeneracy.

{\it Coupling dependence.}
The asymptotes of the thermal conductance show the same temperature dependence for the symmetric mode (both in normal and superradiant phases) and antisymmetric mode.
Meanwhile, we can see a peculiar behavior in its prefactor, depending on $g$, at low and intermediate temperatures.
Particularly, at low temperatures, near the critical point, the thermal conductance scales as $(1-g^2)^{-2}$ on the normal side ($g\to1_-$) and $(g^{4}-1)^{-2}$ on the superradiant side ($g\to1_+$).
Note that, in practice, the thermal conductance does not diverge at the critical point.
Since the excitation energy decreases as $g$ approaches the critical point, the low-temperature condition $k_{\rm B}T\ll\epsilon_{n/s}(g)$ eventually breaks down at a fixed temperature.
Instead, the system enters either the intermediate-temperature regime $k_{\rm B}T\sim\epsilon_{n/s}(g)$ or high-temperature regime $k_{\rm B}T\gg\epsilon_{n/s}(g)$, in which the thermal conductance no longer shows the divergence.

\begin{figure}[tb]
    \centering
    \includegraphics[width=1\linewidth]{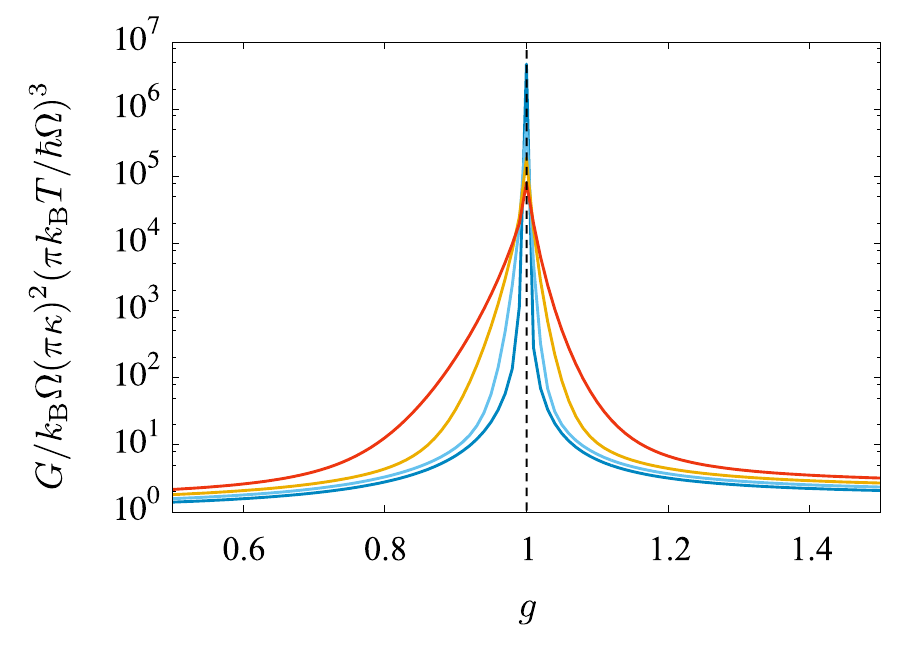}
    \caption{Thermal conductance as a function of $g$ for $\kappa=0.001$ and different temperatures $k_{\rm B}T/\hbar\Omega=0.01$ (blue), 0.02 (light blue), 0.03 (orange), and 0.04 (red).}
    \label{fig:G_g}
\end{figure}

Thermal conductance is plotted as a function of $g$, varying the temperature $k_{\rm B}T/\hbar\Omega=0.01-0.04$ in Fig.~\ref{fig:G_g}.
Here, the thermal conductance increases toward $g=1$ on both the normal and superradiant sides.
As the temperature decreases, its peak becomes sharper around the critical point because the thermal conductance can be described by the low-temperature asymptote, while, for higher temperatures beyond $k_{\rm B}T\ll\epsilon_{n/s}(g)$, the $g$ dependence becomes weaker.
The thermal conductance is not symmetric to $g=1$ and decreases more rapidly on the superradiant side as $g$ deviates from the critical point. 
This asymmetry arises from the fact that the thermal conductance scales as $g^{-4}$ in the superradiant phase, while it scales as $g^2$ in the normal phase (see Eq.~\eqref{eq:Gsp}).

{\it Spin-boson mapping.}
So far, we have considered the low-energy effective Hamiltonian for $\Omega/\Delta\ll1$ and derived the thermal conductance in the temperature range $k_{\rm B}T\ll\hbar\Delta$.
There, thermal excitation between the atomic levels is not taken into account because the atomic part of the Hamiltonian is projected to the lower-energy subspace.
Here, we consider the thermal conductance due to thermal excitation between the atomic levels, which dominates for $k_{\rm B}T\sim\hbar\Delta$.
To this end, we map the quantum Rabi model with Ohmic dissipation to the spin-boson model, $H_{\rm sb}=(\hbar\Delta/2)\sigma_z-\sum_{rk}(\hbar\hat{\lambda}_{k}/2)(\hat{b}_{rk}+\hat{b}_{rk}^\dagger)\sigma_x+\sum_{rk}\hbar\hat{\omega}_{k}\hat{b}_{rk}^\dagger \hat{b}_{rk}$, where $\hat{b}_{rk}$ is the diagonalized bosonic operator~\cite{Fano1961,Kato2007}.
The effect of the single bosonic modes is incorporated into the structured spectral density $\hat{I}(\omega)\equiv\sum_k\hat{\lambda}_{k}^2\delta(\omega-\hat{\omega}_{k})=2\kappa\omega(4\lambda\Omega)^2/[(\omega^2-\Omega^2)^2+(2\pi\kappa\Omega\omega)^2]$.
Assuming a weak coupling $\kappa\ll1$, the reduced density matrix of the atom can be described by the Lindblad equation~\cite{Breuer_text}, $\dot{\rho}(t)=-i(\Delta/2)[\sigma_z,\rho(t)]+\sum_{r,i=\pm}\Gamma_{r,i}[\sigma_i\rho\sigma_i^\dagger-\{\rho(t),\sigma_i^\dagger \sigma_i\}/2]$,
where $\Gamma_{r,+}=\frac{\pi}{2}\hat{I}(\Delta)\hat{n}_r(\Delta)$ and $\Gamma_{r,-}=\frac{\pi}{2}\hat{I}(\Delta)[\hat{n}_r(\Delta)+1]$ are the absorption and emission rates, respectively, and $\hat{n}_r(\omega)$ is the Bose-Einstein distribution function of the diagonalized thermal bath.
Using the Lindblad equation, we obtain the symmetric correlation function, $S(t)=\braket{\{\sigma_x(t),\sigma_x\}}/2$, and then the thermal conductance as~\cite{Yamamoto2021}
\begin{align}
G(T)=\frac{k_{\rm B}}{32}\int d\omega~\hat{I}(\omega)S(\omega)\frac{(\beta\hbar\omega)^2}{\sinh(\beta\hbar\omega)},
\end{align}
where $S(\omega)\approx4\Gamma\Delta^2/[((\omega-\Delta)^2+\Gamma^2)(\omega+\Delta)^2]$ with $\Gamma=(\Gamma_++\Gamma_-)/2$.
For a weak coupling ($\Gamma\ll\Delta$), the symmetric correlation function can be approximated by a delta function, $S(\omega)\approx2\pi\delta(\omega-\Delta)$, which results in the Schottky-type temperature dependence, $G(T,g)\sim k_{\rm B}\Omega\pi\kappa g^2(\hbar\Omega/k_{\rm B}T)^2e^{-\hbar\Delta/k_{\rm B}T}$, for $k_{\rm B}T\lesssim\hbar\Delta$.
Note that the Lindblad approach does not reproduce the thermal conductance for $k_{\rm B}T\ll\hbar\Delta$, and it diminishes as $G(T\ll\hbar\Delta/k_{\rm B})\propto(\Omega/\Delta)^3$ in the classical oscillator limit.
This is attributed to the lack of information on the small energy structure, the order of $\hbar\Omega$, in the Lindblad approach.

{\it Conclusion.}
We investigated the effect of the superradiant phase transition on heat transport through the quantum Rabi model.
Using the Keldysh formalism and perturbation theory, we derived a closed formula for the thermal conductance based on the low-energy effective Hamiltonian for $\Omega/\Delta\ll1$ and asymptotic formulas for the low-temperature, intermediate-temperature, and high-temperature regions.
In particular, at low temperatures, the thermal conductance strongly depends on the coupling $g$.
It rapidly increases toward the critical point, scaling as $G(g)\propto(1-g^2)^{-2}$ on the normal side and $G(g)\propto(g^4-1)^{-2}$ on the superradiant side.
This steep dependence on $g$ would be useful for quantum heat management, such as in quantum heat valves, by tuning $g$ (the coupling strength $\lambda$, the bosonic-mode frequency $\Omega$, or the atomic energy $\Delta$) across the critical point.
While this Letter focused on the classical oscillator limit, it would be interesting to investigate the correction to the thermal conductance due to the finite ratio $\Omega/\Delta$.
In the near future, the superradiant phase transition can be expected to be detected in heat transport experiments, and quantum heat devices could be developed utilizing controllable finite-component quantum phase transitions.

{\it Acknowledgments.}
We thank Takeo Kato and Shunsuke Kamimura for illuminating discussions.
This work was supported by the JST Moonshot R\&D–MILLENNIA
Program Grant No. JPMJMS2061 and JSPS KAKENHI Grant No. 23K03273.

\bibliography{ref}

\end{document}


\title{Supplemental Materials: \\
``Thermal conductance at superradiant phase transition in quantum Rabi model''}

\author{Tsuyoshi Yamamoto}
\email{Contact author: tsuyoshi.yamamoto@cea.fr}
\thanks{\protect\\Present address: Univ. Grenoble Alpes, CEA, Grenoble INP, IRIG, PHELIQS, 38000 Grenoble, France}
\author{Yasuhiro Tokura}
\affiliation{Institute of Pure and Applied Sciences, University of Tsukuba, Tsukuba, Ibaraki 305-8577, Japan}
\date{\today}

\maketitle

\section{Analytical formula for heat current}

Here, we derive the heat current across a bosonic mode-atom-bosonic mode assembly between Ohmic thermal baths by using the Keldysh formalism~\cite{Ojanen2008}.

We consider a two-level atom embedded between single bosonic modes in contact with thermal baths.
The total Hamiltonian is given by
\begin{align}
H&=H_{\rm S}+H_{\rm B}+V, \\
H_{\rm S}
&=\sum_{r=1,2}\hbar\Omega_ra_r^\dagger a_r+\frac{\hbar\Delta}{2}\sigma_z-\sum_{r=1,2}\hbar\lambda_r(a_r+a_r^\dagger)\sigma_x, \\
H_{\rm B}
&=\sum_{r=1,2}H_{{\rm B},r}
=\sum_{r,k}\hbar\omega_{rk}b_{rk}^\dagger b_{rk}, \\
V
&=\sum_{r=1,2}V_r
=\sum_{r,k}\hbar\eta_{rk}(a_r b_{rk}^\dagger+a_r^\dagger b_{rk}),
\end{align}
where we have assumed the rotating-wave approximation.
The parameters and the operators are explained in the main text.
We refer to the Hamiltonian without the coupling between the bosonic modes and the thermal baths $V$ as the unperturbed Hamiltonian $H_0=H_{\rm S}+H_{\rm B}$.

The heat current flowing out of thermal bath 1 is defined as
\begin{align}
J_1(t)
\equiv-\Braket{\frac{dH_{{\rm B},1}(t)}{dt}}
=-\frac{1}{i\hbar}\Braket{[H_{{\rm B},1}(t),H(t)]}
=i\hbar\sum_{k}\omega_{1k}\eta_{1k}\braket{a_1(t)b_{1k}^\dagger(t)-a_1^\dagger(t)b_{1k}(t)}
\end{align}
where the quantum mechanical average $\braket{\mathcal{O}}$ and time evolution $\mathcal{O}(t)$ are with respect to the total Hamiltonian.
Introducing the lesser Green functions,
\begin{align}
G^<_{A,B}(t,t')=\frac{1}{i\hbar}\braket{B(t')A(t)},
\end{align}
the heat current can be expressed as
\begin{align}
J_1(t)=-2\hbar^2\sum_{k}\omega_{1k}\eta_{1k}{\rm Re}\left[G^<_{a_1,b_{1k}^\dagger}(t_1,t_2)\right]_{t=t_1=t_2}.
\end{align}

We calculate the lesser Green function by using the Keldysh formalism.
To this end, we first consider the Keldysh Green function,
\begin{align}
G_{a_1,b_{1k}^\dagger}(\tau_1,\tau_2)
=\frac{1}{i\hbar}\braket{T_Ka_1(\tau_1)b_{1k}^\dagger(\tau_2)},
\end{align}
where $\tau_{1,2}$ is the time on the Keldysh contour and  $T_K$ is its time-ordering operator.
Using the standard technique of the Keldysh Green function, we can rewrite the Keldysh Green function as
\begin{align}
G_{a_1,b_{1k}^\dagger}(\tau_1,\tau_2)
=\frac{1}{i\hbar}\braket{T_KS_K\tilde{a}_1(\tau_1)\tilde{b}_{1k}^\dagger(\tau_2)}_0,
\end{align}
where $\braket{\mathcal{O}}_0$ is the quantum mechanical average for the unperturbed Hamiltonian $H_0$ and $\tilde{\mathcal{O}}(\tau)=e^{iH_0\tau/\hbar}\mathcal{O}e^{-iH_0\tau/\hbar}$ is the time evolution in the interaction picture.
Here, the Keldysh action $S_K$ is
\begin{align}
S_K=\sum_{n=0}^\infty \left(\frac{1}{i\hbar}\right)^n\frac{1}{n!}\int_K d\xi_1\cdots \int_K d\xi_n~\tilde{V}(\xi_1)\cdots \tilde{V}(\xi_n),
\end{align}
where $\xi_i$ is the time on the Keldysh contour and $\int_K d\xi$ represents the integration over the Keldysh contour.
Therefore, the Keldysh Green function can be expanded as
\begin{align}
G_{a_1,b_{1k}^\dagger}(\tau_1,\tau_2)
&=\frac{1}{i\hbar}\sum_{n=0}^\infty \left(\frac{1}{i\hbar}\right)^n\frac{1}{n!}\int_K d\xi_1\cdots\int_K d\xi_j\cdots\int_K d\xi_n\braket{T_K\tilde{V}(\xi_1)\cdots \tilde{V}(\xi_j)\cdots\tilde{V}(\xi_n)\tilde{a}_1(\tau_1)\tilde{b}_{1k}^\dagger(\tau_2)}_0 \nonumber\\
&=\frac{1}{i\hbar}\sum_{rq}\hbar\eta_{rq}\sum_{n=0}^\infty \left(\frac{1}{i\hbar}\right)^n\frac{1}{n!}\int_K d\xi_1\cdots\int_K d\xi_j\cdots\int_K d\xi_n \nonumber\\
&\qquad\qquad\qquad\qquad\times
\braket{T_K\tilde{V}(\xi_1)\cdots [\tilde{a}_r(\xi_j)\tilde{b}_{rq}^\dagger(\xi_j)+\tilde{a}_r^\dagger(\xi_j)\tilde{b}_{rq}(\xi_j)]\cdots\tilde{V}(\xi_n)\tilde{a}_1(\tau_1)\tilde{b}_{1k}^\dagger(\tau_2)}_0 \nonumber\\
&=\frac{1}{i\hbar}\sum_{rq}\hbar\eta_{rq}\sum_{n=0}^\infty \left(\frac{1}{i\hbar}\right)^n\frac{1}{n!}\int_K d\xi_1\cdots\int_K d\xi_j\cdots\int_K d\xi_n \nonumber\\
&\qquad\qquad\qquad\qquad\times
\braket{T_K\tilde{V}(\xi_1)\cdots\tilde{V}(\xi_n)\tilde{a}_r(\tau_1)\tilde{a}_1^\dagger(\xi_j)}_0
\braket{T_K\tilde{b}_{rq}(\xi_j)\tilde{b}_{1k}^\dagger(\tau_2)}_0 \times n \nonumber\\
&=\sum_{rq}\hbar\eta_{rq}\int_K d\xi_j~\frac{1}{i\hbar}\braket{T_KS_K\tilde{a}_1(\tau_1)\tilde{a}_r^\dagger(\xi_j)}_0\times\frac{1}{i\hbar}\braket{T_K\tilde{b}_{rq}(\xi_j)\tilde{b}_{1k}^\dagger(\tau_2)}_0 \nonumber\\
&=\hbar\eta_{1k}\int_K d\xi_j~G_{a_1,a_1^\dagger}(\tau_1,\xi_j)g_{1k}(\xi_j,\tau_2),
\end{align}
where we have used the Wick theorem and $\braket{T_K\tilde{b}_{rq}(\xi_j)\tilde{b}_{1k}^\dagger(\tau_2)}_0=\braket{T_K\tilde{b}_{1k}(\xi_j)\tilde{b}_{1k}^\dagger(\tau_2)}_0\delta_{r1}\delta_{kq}$.
Here, $g_{1,k}(\tau,\tau')$ is the Keldysh Green function of the non-interacting bosonic field in the thermal bath,
\begin{align}
g_{1k}(\tau,\tau')
=\frac{1}{i\hbar}\braket{T_K\tilde{b}_{1k}(\tau)\tilde{b}_{1k}^\dagger(\tau')}_0.
\end{align}
Using the Langreth rule~\cite{Stefanucci_text}, we obtain the lesser Green function,
\begin{align}
G^<_{a_1,b_{1k}^\dagger}(t_1,t_2)
=\hbar\eta_{1k} \int ds~\left[G^{\rm R}_{a_1,a_1^\dagger}(t_1,s)g^<_{1k}(s,t_2)+G^<_{a_1,a_1^\dagger}(t_1,s)g^{\rm A}_{1k}(s,t_2)\right].
\end{align}
Here, the retarded Green function is
\begin{align}
G_{A,B}^{\rm R}(t,t')=\frac{1}{i\hbar}\theta(t-t')\braket{[A(t),B(t')]},
\end{align}
and the lesser and advanced Green functions of the non-interacting bosonic field are
\begin{align}
g^<_{1k}(t,t')
&=\frac{1}{i\hbar}\braket{\tilde{b}_{1k}^\dagger(t')\tilde{b}_{1k}(t)}_0
=\frac{1}{i\hbar}e^{-i\omega_{1k}(t-t')}n_1(\omega_k) \\
g^{\rm A}_{1k}(t,t')
&=-\frac{1}{i\hbar}\theta(t'-t)\braket{[\tilde{b}_{1k}(t),\tilde{b}_{1k}^\dagger(t')]}_0
=-\frac{1}{i\hbar}\theta(t'-t)e^{-i\omega_{1k}(t-t')},
\end{align}
respectively, where $n_r(\omega_k)=1/(e^{\beta_r\hbar\omega_k}-1)$ is the Bose-Einstein distribution of the heat bath $r$ with the temperature $T_r=1/(\beta_r k_{\rm B})$.
Since the Green functions depend on only the time difference, the lesser Green function reads, in terms of the Fourier expression, as
\begin{align}
G^<_{a_1,b_{1k}^\dagger}(t_1,t_2)
=\hbar\eta_{1k}\int \frac{d\omega}{2\pi}e^{-i\omega(t_1-t_2)}\left[G^{\rm R}_{a_1,a_1^\dagger}(\omega)g^<_{1k}(\omega)+G^<_{a_1,a_1^\dagger}(\omega)g^{\rm A}_{1k}(\omega)\right].
\end{align}

As a result, we obtain the heat current,
\begin{align}
J_1(t)
&=-2\hbar^2\sum_{k}\omega_{1k}\eta_{1k}{\rm Re}\left[\hbar\eta_{1k}\int \frac{d\omega}{2\pi}e^{-i\omega(t_1-t_2)}\left[G^{\rm R}_{a_1,a_1^\dagger}(\omega)g^<_{1k}(\omega)+G^<_{a_1,a_1^\dagger}(\omega)g^{\rm A}_{1k}(\omega)\right]\right]_{t=t_1=t_2} \nonumber\\
&=-2\sum_{k}\hbar\omega_{1k}(\hbar\eta_{1k})^2\int \frac{d\omega}{2\pi}~{\rm Re}\left[G^{\rm R}_{a_1,a_1^\dagger}(\omega)g^<_{1k}(\omega)+G^<_{a_1,a_1^\dagger}(\omega)g^{\rm A}_{1k}(\omega)\right].
\end{align}
The summations about $k$ are rewritten as
\begin{align}
\sum_k\hbar\omega_{1,k}(\hbar\eta_{1k})^2g_{1k}^<(\omega)
&=-2i\pi\hbar^2\omega I_1(\omega)n_1(\omega), \\
\sum_k\hbar\omega_{1,k}(\hbar\eta_{1k})^2g_{1k}^{\rm A}(\omega)
&=\sum_k \hbar^2\omega_{1k}\frac{\eta_{1k}^2}{\omega-\omega_{1k}-i\delta}
=\sum_k\hbar^2\omega_{1k}\eta_{1k}^2\mathcal{P}\frac{1}{\omega-\omega_{1k}}+i\pi\hbar^2\omega I_1(\omega),
\end{align}
where $I_r(\omega)\equiv\sum_{k}\eta_{rk}^2\delta(\omega-\omega_{rk})$ is the spectral density, characterizing the properties of the thermal bath.
Now, assuming the symmetric case, $I(\omega)=I_1(\omega)=I_2(\omega)$, the heat current reads as
\begin{align}
J_1
=-2\hbar\int d\omega~\hbar\omega I(\omega)\left\{{\rm Im}\left[G^{\rm R}_1(\omega)\right]n_1(\omega)
-\frac{1}{2}{\rm Im}\left[G^<_1(\omega)\right]\right\},
\end{align}
where $G_{A,A^\dagger}^<(\omega)$ is pure imaginary and we have used the notation, $G_{a_1,a_1^\dagger}^{{\rm R},<}(\omega)=G_{1}^{{\rm R},<}(\omega)$, in the main text.
In the steady-state limit, where $J=J_1=-J_2=(J_1-J_2)/2$, we obtain Eq.~(2) in the main text for the heat current.

\section{Perturbation theory in $V$}
\label{sm:pertubation}

Here, we perform perturbation theory on the interaction between the single bosonic mode and the thermal bath to express the retarded Green function in the thermal conductance in terms of isolated system thermal baths.
For the perturbative approach, we divide the total Hamiltonian $H=H_++H_-$, where
\begin{align}
H_\pm
&=H_{{\rm S},\pm}+H_{{\rm B},\pm}+V_{\pm}, \\
H_{{\rm S},+}
&=\hbar\Omega a_+^\dagger a_++\frac{\hbar\Delta}{2}\sigma_z-\sqrt{2}\hbar\lambda(a_++a_+^\dagger)\sigma_x, \\
H_{{\rm S},-}
&=\hbar\Omega a_-^\dagger a_-, \\
H_{{\rm B},\pm}
&=\sum_k\hbar\omega_{k}b_{\pm,k}^\dagger b_{\pm,k}, \\
V_\pm
&=\sum_k\hbar\eta_k(a_\pm b_{\pm,k}^\dagger+a_\pm^\dagger b_{\pm,k}),
\end{align}
into the unperturbed Hamiltonian, $H_{0,\pm}=H_{{\rm S},\pm}+H_{{\rm B},\pm}$, and the interaction Hamiltonian, $V=V_++V_-$.

First, we introduce the Matsubara Green function,
\begin{align}
\mathcal{G}_\pm(\tau)=-\braket{T_\tau a_\pm(\tau)a_\pm^\dagger(0)}_\pm,
\end{align}
where $\mathcal{O}_\pm(\tau)=e^{\tau H_\pm}\mathcal{O}e^{-\tau H_\pm}$,  $T_\tau$ is the imaginary-time ordering (not time on the Keldysh contour here), and $\braket{\mathcal{O}}_\pm$ denotes the quantum mechanical average for the total Hamiltonian $H_\pm$.
When expressing the retarded Green function and the Matsubara Green function in the Lehman representation, we have
\begin{align}
G^{\rm R}_\pm(\omega)
&=\int_{-\infty}^\infty dt~e^{i\omega t}G^{\rm R}_\pm(t)
=\frac{1}{Z_\pm}\sum_{n,m}|\braket{n_\pm|a_\pm|m_\pm}|^2\frac{e^{-\beta E_n^\pm}-e^{-\beta E_m^\pm}}{\hbar\omega+E_n^\pm-E_m^\pm+i\delta}, \\
\mathcal{G}_\pm(i\xi_n)
&=\int_0^\beta d\tau~e^{i\xi_n \tau}\mathcal{G}_\pm(\tau)
=\frac{1}{Z_\pm}\sum_{n,m}|\braket{n_\pm|a_\pm|m_\pm}|^2\frac{e^{-\beta E_n^\pm}-e^{-\beta E_m^\pm}}{i\xi_n+E_n^\pm-E_m^\pm},
\end{align}
respectively, where $E_n^\pm$ is the eigenenergy and $\ket{n_\pm}$ is the eigenstate of the total Hamiltonian, i.e., $H_\pm\ket{n_\pm}=E_n^\pm\ket{n_\pm}$ and $Z_\pm$ is the partition function.
Here, $\xi_n=2\pi n/\beta=2\pi n k_{\rm B}T$ is the Matsubara frequency.
Comparing these expressions, the retarded Green function is obtained from the Matsubara Green function by analytical continuation,
\begin{align}
G^{\rm R}_\pm(\omega)=\mathcal{G}_\pm(i\xi_n\to\hbar\omega+i\delta).
\end{align}

Using the imaginary-time evolution operator,
\begin{align}
\mathcal{U}_\pm(\beta)
&=\sum_{n=0}^\infty \frac{(-1)^n}{n!}\int_0^\beta d\tau_1 \cdots \int_0^\beta d\tau_n~T_\tau \tilde{V}_\pm(\tau_1)\cdots \tilde{V}_\pm(\tau_n),
\end{align}
the Matsubara Green function can be expressed, in the interaction picture, as
\begin{align}
\mathcal{G}_\pm(\tau,\tau')
=-\frac{\braket{T_\tau\mathcal{U}_\pm(\beta)\tilde{a}_\pm(\tau)\tilde{a}_\pm^\dagger(\tau')}_{0,\pm}}{\braket{\mathcal{U}_\pm(\beta)}_{0,\pm}},
\end{align}
where $\tilde{\mathcal{O}}_\pm(\tau)=e^{\tau H_{0,\pm}}\mathcal{O}_\pm e^{-\tau H_{0,\pm}}$ and $\braket{\mathcal{O}}_{0,\pm}$ is the quantum mechanical average with the unperturbed Hamiltonian $H_{0,\pm}$.
The standard field theory provides~\cite{Mahan_text}
\begin{align}
\mathcal{G}_\pm(\tau,\tau')
&=-\sum_{n=0}^\infty\frac{(-1)^n}{n!}\int_0^\beta d\tau_1\cdots\int_0^\beta d\tau_n\frac{\braket{T_\tau \tilde{V}_\pm(\tau_1)\cdots \tilde{V}_\pm(\tau_n)\tilde{a}_\pm(\tau)\tilde{a}_\pm^\dagger(\tau')}_{0,\pm}}{\braket{\mathcal{U}_\pm(\beta)}_{0,\pm}} \nonumber\\
&=\mathcal{G}_{0,\pm}(\tau,\tau')+\int_0^\beta d\tau_1 \int_0^\beta d\tau_2~\mathcal{G}_{0,\pm}(\tau,\tau_2)\Sigma_\pm(\tau_2,\tau_1)\mathcal{G}_{0,\pm}(\tau_1,\tau')+\cdots,
\end{align}
where we have used $\braket{\tilde{V}^{2n+1}_\pm(\tau)}_{0,\pm}=0$.
Here, $\mathcal{G}_{0,+}(\tau,\tau')$ is the Matsubara Green function of the quantum Rabi model and $\mathcal{G}_{0,-}(\tau,\tau')$ is that of the single resonator,
\begin{align}
\mathcal{G}_{0,\pm}(\tau,\tau')
=\mathcal{G}_{0,\pm}(\tau-\tau')
=-\braket{T_\tau \tilde{a}_\pm(\tau-\tau')\tilde{a}_\pm^\dagger(0)}_{0,\pm},
\end{align}
and $\Sigma_\pm(\tau,\tau')$ is the self-energy,
\begin{align}
\Sigma_\pm(\tau,\tau')
=\Sigma_\pm(\tau-\tau')
=\sum_k(\hbar\eta_k)^2 g_{\pm k}(\tau-\tau').
\end{align}
where $g_{\pm k}(\tau-\tau')=-\braket{T_\tau \tilde{b}_{\pm k}(\tau-\tau')\tilde{b}_{\pm k}^\dagger(0)}_{0,\pm}$ is the Matsubara Green function for a non-interacting bosonic field.
In Fourier space, we obtain the Dyson equation,
\begin{align}
\mathcal{G}_\pm(i\xi_n)
=\frac{\mathcal{G}_{0,\pm}(i\xi_n)}{1-\mathcal{G}_{0,\pm}(i\xi_n)\Sigma_\pm(i\xi_n)},
\end{align}
where the Fourier transformation of the self-energy is given as
\begin{align}
\Sigma_\pm(i\xi_n)
=\sum_k(\hbar\eta_k)^2\frac{1}{i\xi_n-\hbar\omega_k}.
\end{align}

Finally, the analytic continuation provides the retarded Green function as
\begin{align}
G^{\rm R}_\pm(\omega)
=\frac{g^{{\rm R}}_\pm(\omega)}{1-g^{{\rm R}}_\pm(\omega)\Sigma_\pm(i\xi_n\to\hbar\omega+i\delta)},
\end{align}
where $g^{{\rm R}}_+(\omega)$ is the Fourier transformation of the retarded Green function of the quantum Rabi model and $g^{{\rm R}}_-(\omega)$ is that of the single bosonic mode.
The self-energy in real frequency space is written as
\begin{align}
\Sigma(\omega)=\Sigma_\pm(i\xi_n\to\hbar\omega+i\delta)
=\sum_{k}(\hbar\eta_{k})^2\frac{1}{\hbar\omega-\hbar\omega_{k}+i\delta}
=\hbar\sum_{k}\eta_{k}^2\frac{\mathcal{P}}{\omega-\omega_{k}}-i\pi\hbar I(\omega).
\end{align}

\section{Asymmetric contribution to heat current}

Here, we consider another contribution to the thermal conductance $\bar{G}(T)=\lim_{\Delta T\to0}\bar{J}/\Delta T$ different from Eq.~(3) in the main text.
It arises from the heat current $\bar{J}$ due to the asymmetry of the resonator,
\begin{align}
\bar{J}
=-\frac{\hbar}{2}\int d\omega~\hbar\omega I(\omega)\left\{{\rm Im}\left[G^{\rm R}_1(\omega)-G^{\rm R}_2(\omega)\right]n(\omega,T)
-\frac{1}{2}{\rm Im}\left[G^<_1(\omega)-G^<_2(\omega)\right]\right\}.
\end{align}
The retarded Green functions can be expressed in terms of the symmetric and antisymmetric modes as
\begin{align}
G^{\rm R}_{1/2}(\omega)
=\frac{G^{\rm R}_{+}(\omega)+G^{\rm R}_{-}(\omega)}{2}.
\end{align}
Note that, because the symmetric and antisymmetric modes are decoupled from each other, $[H_+, H_-]=0$, the retarded Green functions of the different modes vanish,
\begin{align}
G^{\rm R}_{a_\pm,a_\mp^\dagger}(t)
&=\frac{1}{i\hbar}\theta(t)\braket{[e^{i(H_++H_-)t/\hbar}a_\pm e^{-i(H_++H_-)t/\hbar},a_\mp^\dagger]} \nonumber\\
&=\frac{1}{i\hbar}\theta(t)\braket{[e^{iH_\pm t/\hbar}a_\pm e^{-iH_\pm t/\hbar},a_\mp^\dagger]} \nonumber\\
&=\frac{1}{i\hbar}\theta(t)\braket{e^{iH_\pm t/\hbar}[a_\pm,a_\mp^\dagger]e^{-iH_\pm t/\hbar}}
=0.
\end{align}
Since the low-energy effective Hamiltonian is quadratic, the retarded Green function is independent of the temperature, as discussed in the main text.
Therefore, the retarded components in $\bar{J}$ vanish.

The lesser components in $\bar{J}$ are
\begin{align}
G^<_{1}(t,t')-G^<_{2}(t,t')
=\frac{1}{i\hbar}\left[\braket{a_+^\dagger(t')}\braket{a_-(t)}+\braket{a_-^\dagger(t')}\braket{a_+(t)}\right].
\end{align}
Using perturbation theory on the interaction between the bosonic modes and the thermal baths in a similar way to SM.~\ref{sm:pertubation}, the coherence of the bosonic mode can be expressed as 
\begin{align}
\braket{a_\pm(t)}
=\frac{\braket{TU_\pm(\infty,-\infty)\tilde{a}_\pm(t)}_{0,\pm}}{\braket{U_\pm(\infty,-\infty)}_{0,\pm}}
\end{align}
where $T$ is the time-ordering operator and $U_\pm(t,t_0)$ is the time-evolution operator,
\begin{align}
U_\pm(t,t_0)=1+\sum_{n=1}\frac{1}{n!}\left(\frac{1}{i\hbar}\right)^n \int_{t_0}^t dt_1\cdots \int_{t_0}^t dt_n~T\tilde{V}_\pm(t_1)\cdots\tilde{V}_\pm(t_n).
\end{align}
Only the leading order remains, $\braket{a_\pm(t)}\approx\braket{\tilde{a}_\pm(t)}_{0,\pm}$ because $\braket{\tilde{b}_{\pm,k}^{2n+1}}_{0,\pm}=0$.
For the symmetric mode, the coherence remains finite in the superradiant phase.
However, since $\braket{a_-(t)}=0$ for the antisymmetric mode, the lesser components in $\bar{J}$ also vanish.

Therefore, the heat current $\bar{J}$ due to the asymmetry of the bosonic modes does not contribute to the thermal conductance, and hence we obtain Eq.~(3) in the main text for the thermal conductance.

\bibliography{ref}